\documentclass[%
 reprint,
 amsmath,amssymb,
 aps,
]{revtex4-2}

\usepackage{graphicx}
\usepackage{dcolumn}
\usepackage{bm}

\usepackage{blindtext}
\usepackage{subcaption}
\usepackage{siunitx}

\newcommand{\muSR}{$\mathrm{\mu}$SR}
\newcommand{\LEmuSR}{LE-$\mathrm{\mu}$SR}

\begin{document}

\preprint{APS/123-QED}

\title{Improving the low-energy muon beam quality of the LEM beamline at PSI: \\ Characterisation of ultra-thin carbon foils}

\author{Gianluca Janka}
\email{gianluca.janka@psi.ch}
\affiliation{Laboratory for Muon Spin Spectroscopy, Paul Scherrer Institute, CH-5232 Villigen PSI, Switzerland}

\author{Maria Mendes Martins}
\affiliation{Laboratory for Muon Spin Spectroscopy, Paul Scherrer Institute, CH-5232 Villigen PSI, Switzerland}
\affiliation{Advanced Power Semiconductor Laboratory, ETH Zürich, Physikstrasse 3, CH-8092 Zürich,  Switzerland}

\author{Xiaojie Ni}
\affiliation{Laboratory for Muon Spin Spectroscopy, Paul Scherrer Institute, CH-5232 Villigen PSI, Switzerland}
\affiliation{State Key Laboratory of Particle Detection and Electronics, University of Science and Technology of China, Hefei 230026, China}

\author{Zaher Salman}
\affiliation{Laboratory for Muon Spin Spectroscopy, Paul Scherrer Institute, CH-5232 Villigen PSI, Switzerland}

\author{Andreas Suter}
\affiliation{Laboratory for Muon Spin Spectroscopy, Paul Scherrer Institute, CH-5232 Villigen PSI, Switzerland}

\author{Thomas Prokscha}
\email{thomas.prokscha@psi.ch}
\affiliation{Laboratory for Muon Spin Spectroscopy, Paul Scherrer Institute, CH-5232 Villigen PSI, Switzerland}
\date{\today}

\begin{abstract}
The Low-Energy Muon beamline (LEM) at the Paul Scherrer Institute currently stands as the world's only facility providing a continuous beam of low-energy muons with keV energies for conducting muon spin rotation experiments on a nanometer depth scale in heterostructures and near a sample's surface. As such, optimizing the beam quality to reach its full potential is of paramount importance. 
One of the ongoing efforts is dedicated to improving the already applied technique of single muon tagging through the detection of secondary electrons emerging from an ultra-thin carbon foil. 
In this work, we present the results from installing a thinner foil with a nominal thickness of \SI{0.5}{\micro\gram\per\centi\meter\squared} and compare its performance to that of the previously installed foil with a nominal thickness of \SI{2.0}{\micro\gram\per\centi\meter\squared}. Our findings indicate improved beam quality, characterized by smaller beam spots, reduced energy loss and straggling of the muons, and enhanced tagging efficiency. Additionally, we introduce a method utilizing blue laser irradiation for cleaning the carbon foil, further improving and maintaining its characteristics.\end{abstract}
\maketitle


\section{\label{sec:Intro}Introduction}

Positively charged muons ($\mu^+$) serve as sensitive probes to study magnetic and electronic properties of materials using the muon spin rotation technique (\muSR) in condensed matter and material science \cite{2022_Hillier, 2022_Blundell, 2024_Amato}. At the Swiss Muon Source (S$\mu$S) located at the Paul Scherrer Institute (PSI) in Switzerland, such measurements typically utilize a continuous beam of surface muons with an energy of \SI{4}{\mega\electronvolt}. At such high kinetic energies, the \muSR\ measurements are restricted to studying the bulk part of the material and cannot give any information about the magnetic properties near the surface. To access this region, the low-energy muon beamline (LEM) was established at the $\mu$E4 beamline of S$\mu$S \cite{2000_Morenzoni,2008_Prokscha}, delivering the world's most powerful 4-MeV muon beam to the LEM facility. 

In the case of LEM, surface muons traverse a solid rare gas layer, where a fraction of them is moderated to around \SI{15}{\electronvolt} and is subsequently re-accelerated up to \SI{20}{\kilo\electronvolt} \cite{1994_Morenzoni,2004_Morenzoni,2001_Prokscha}. At these kinetic energies, the muons can be guided and focused by electrostatic means. By applying a bias to the samples under investigation, the muons can be decelerated or accelerated to achieve the desired implantation energy between \SI{1}{\kilo\electronvolt} and \SI{30}{\kilo\electronvolt}. This approach not only allows to probe the surface region, but also enables depth-dependent measurements in the range of a few nanometers up to about \SI{300}{\nano\meter}, and direct comparison with bulk \muSR\ measurements. 
However, conducting experiments with a low-energy muon beam introduces a range of new challenges. Due to the moderation efficiency being around $10^{-5}$ to $10^{-4}$ \cite{2001_Prokscha, 2004_Morenzoni}, the low-energy
muon (LE-$\mu^+$) rate at the moderator is \SI{15(25)}{\kilo\hertz} for a solid Ar(Ne) moderator at a proton beam current of \SI{2.0}{\milli\ampere}. The
low-energy \muSR\ (\LEmuSR) event rates are currently limited to a maximum of approximately \SI{2.5}{\kilo\hertz}, where the reduction to the initial LE-$\mu^+$ is due to transport and detector efficiencies. Therefore, there is an ongoing, dedicated effort to enhance the beamline's performance with the goal of increasing the usable muon rate and improving beam quality. Recent advancements include the study of an upgrade to the muon beamline $\mu$E4, which is expected to boost the muon rate by approximately a factor of 3 arriving at the LEM moderator \cite{2022_Zhou}. Additionally, the introduction of a collimator in the beamline now comfortably allows measurements of sample sizes down to $10\times10$\si{\milli\meter\squared}, opening up the possibility of mounting multiple samples on the same sample plate \cite{2023_Ni}.

\begin{figure*}[t!]
    \centering
    \includegraphics[width=0.99\textwidth, trim={35 60 60 60},clip]{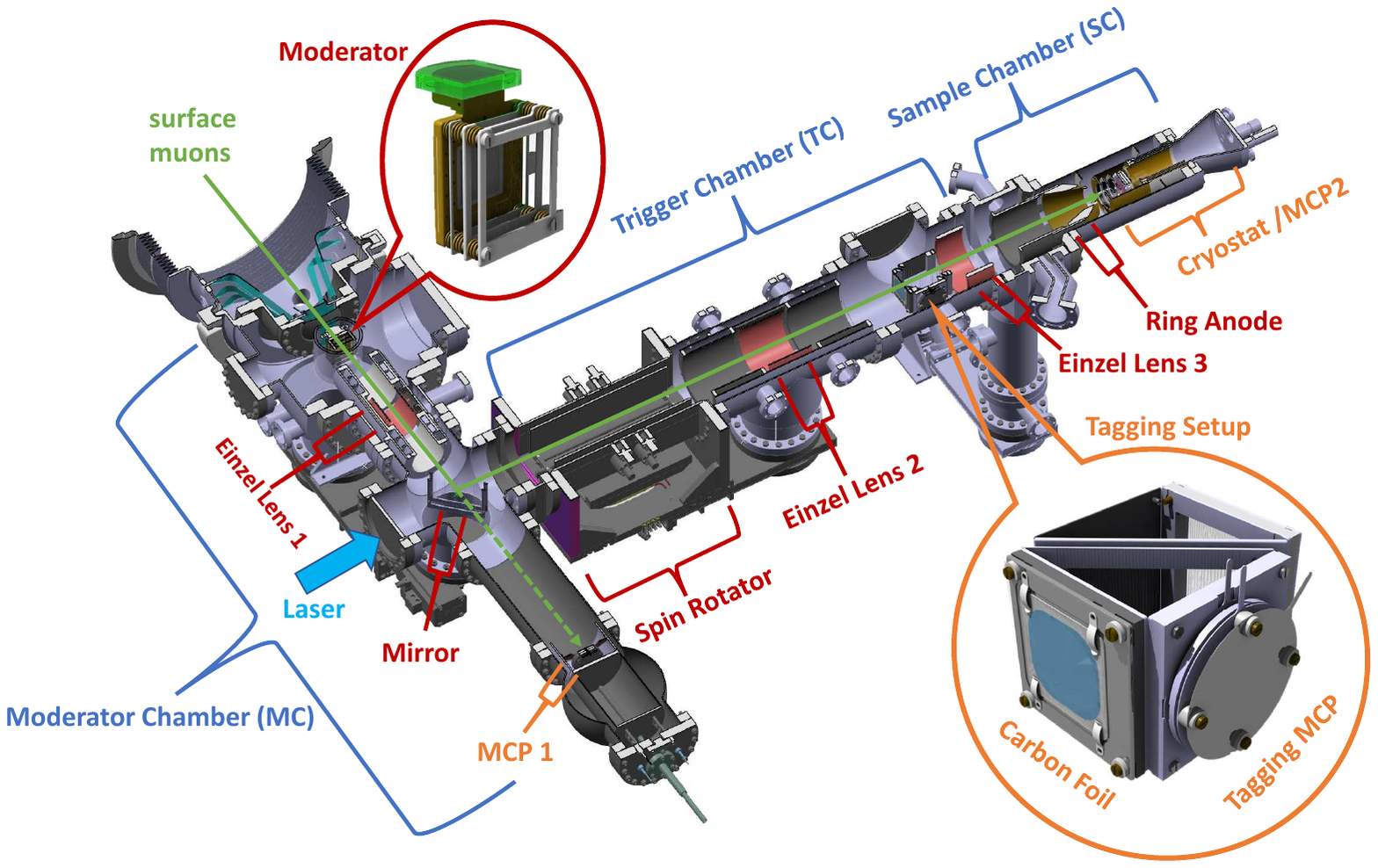}
	  \caption[]{\label{fig:lem}Cross-section view of the LEM beamline CAD model. The three different chambers are labelled in blue, the electrostatic and magnetic elements to transport the muons in red and the detection setups in orange. The moderator and the tagging setup are enlarged for better visualization.}
\end{figure*}

One of the key, but also one of the currently limiting elements of the \LEmuSR\ measurements at LEM, is the ultra-thin carbon foil used for single muon tagging \cite{1998_Hofer, 2015_Khaw}. Due to the continuous nature of the beam, it is crucial to indirectly detect each incoming muon as efficiently as possible to determine the starting time of a valid \muSR\ event. When a $\mu^+$ passes through the carbon foil, it interacts with the material, losing a portion of its kinetic energy. During this process, it also releases a few secondary electrons from the surface of the foil, which are detected by a nearby microchannel plate (MCP), providing the start signal. 

The necessity of placing an ultra-thin carbon foil in the path of the $\mu^+$ beam for tagging introduces certain limitations in the \LEmuSR\ measurements. Interactions with the foil cause the $\mu^+$ particles to scatter, significantly impacting the achievable beam spot size. This limitation restricts the minimum sample size that can be effectively measured without compromising statistics or acquiring more background noise. Additionally, the beam loses its mono-energetic nature and carries an energy distribution, which, depending on the foil characteristics, may not be symmetric. Consequently, the energy distribution introduces an uncertainty in the arrival time of the $\mu^+$ particles at the sample, which in turn leads to the washing out of high-frequency spin precessions and therefore setting an upper boundary of frequencies that can be resolved.

To mitigate these limitations, it is advantageous to use thinner carbon foils, reducing the energy loss and distribution width (straggling) of the $\mu^+$, while maintaining a high tagging efficiency by still having a large, homogeneous foil resting on a highly transparent grid. In 2018, a foil with a thickness of roughly \SI{10}{\nano\meter}, equivalent to \SI{2.0}{\micro\gram\per\centi\meter\squared}, was transferred in-house onto a frame with copper mesh (45 lines per inch, \SI{20}{\micro\meter} wires, \SI{93}{\percent} transparency, circular area with diameter of \SI{40}{\milli\meter}, ordered from ``Precision Eforming" \cite{EForming}) and installed in LEM. 

In this work, we present the results from a newly installed ultra-thin carbon foil with a nominal thickness of approximately \SI{2}{\nano\meter} (equivalent to roughly \SI{0.5}{\micro\gram\per\centi\meter\squared}), purchased from ``The Arizona Carbon Foil Company (ACF)" \cite{ACF}. The foil was mounted on a grid with the same characteristics and dimensions as that previously used. We compare these results with data collected in 2020 using the thick foil, examining parameters such as energy loss, straggling, tagging efficiency, neutrals formation efficiency, and achievable beam spots for both protons and $\mu^+$s. Additionally, we present a method based on illuminating the foil with a blue laser to clean the foil, improve its characteristics, and prolong optimal operation even further.

\section{\label{sec:Exp}Experimental Setup}

The vacuum system of the LEM beamline (Fig.~\ref{fig:lem}) consists of three chambers. 
\begin{enumerate}
    \item Moderator Chamber (MC), where the surface muons are moderated and re-accelerated to a kinetic energy of up to \SI{20}{\kilo\electronvolt}. The low-energy muons beam is then deflected by 90\si{\degree} using an electrostatic mirror, while the unmoderated muons continue on a straight path, and are stopped and detected by an MCP (MCP1). 
    \item Trigger Chamber (TC), where muons first pass through a spin rotator \cite{2012_Salman}. By tuning the ratio of orthogonal E and B fields, only the low-energy muons are allowed to pass through without deflection, while all other beam contamination with the same kinetic energy but different velocity is rejected. Subsequently, the low-energy muons are detected using the ultra-thin carbon foil and a tagging MCP, providing the start time of a \muSR\ event. The foil is biased at \SI{-3.48}{\kilo\volt} to accelerate the $\mu^+$ towards the foil. Upon leaving the foil, the $\mu^+$ has a chance to capture an electron, forming neutral muonium atoms. 
    \item Sample Chamber (SC), where the sample of interest is mounted, typically on a cold-finger cryostat. To maximize the number of muons stopping in the sample, they are focused beforehand using a segmented ring anode \cite{2017_Xiao}. The SC is mounted in the bore of a magnet to apply magnetic fields on the studied sample. Positron counter detectors which are read out by avalanche photodiodes (APDs) are used to detect the muon decay and its spin direction, which acts as the end time of the \muSR\ event. Alternatively, a position-sensitive MCP (MCP2) can be installed at the sample position to monitor the beam spot and rates.
\end{enumerate}

In the MC, a tantalum wire is installed close to the moderator, which can be heated by an electric current to release electrons. We speculate that these electrons bombard the biased moderator, dissociating adsorbed gases or molecules (e.g. H$_{2}$, H$_{2}$O), and freeing protons that can then be accelerated from the moderator, forming a low energy proton beam. This allows for testing most characteristics of the LEM beamline with protons when $\mu^+$ are not available.


\section{\label{sec:foil}Foil Characterisation}

\begin{figure}[t!]    \centering
	    \includegraphics[width=1\columnwidth, trim={10 5 35 25},clip]{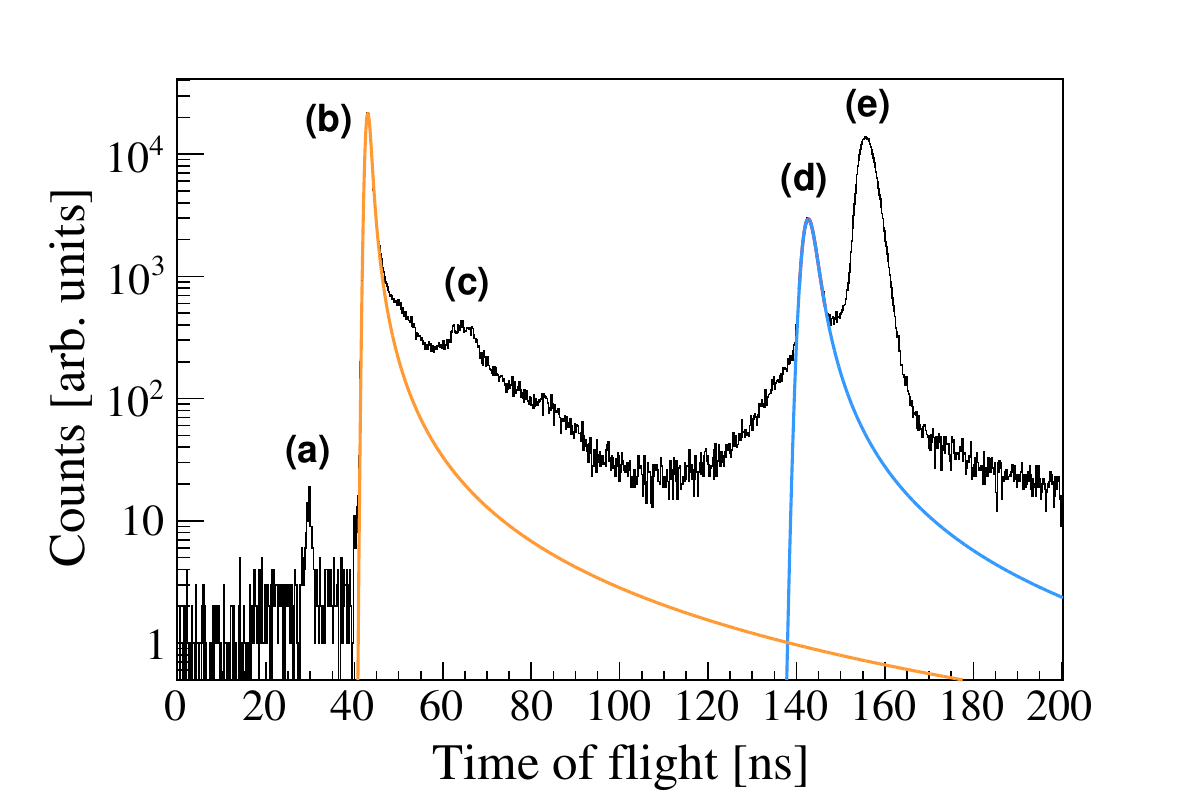}
	       \caption[]{\label{fig:M_TOF_12keV}A typical time-of-flight spectrum, recorded with muons accelerated to \SI{12}{\kilo\electronvolt} at the moderator. Several peaks are visible in the spectrum and described in the text in more detail. The prompt peak at (b) (orange) as well as the muonium peak at (d) (blue) are fitted with a Landau function convoluted with a Gaussian to determine the most probable arrival time.} 
\end{figure}

To characterize the foil, time-of-flight (TOF) spectra were collected between the tagging timestamp and the particle's detection on the position-sensitive MCP2. Due to the biased carbon foil, the neutral atoms formed continue their path with an additional energy of \SI{+3.48}{\kilo\electronvolt} compared to the charged incident particles, resulting in a time separation in the TOF spectrum. To ensure consistent transport efficiency from the foil to the MCP2, all electric fields after the tagging setup were turned off. A typical TOF spectrum is shown in Fig.~\ref{fig:M_TOF_12keV}, where the separation between neutral muonium (peak (d)) and muons (peak (e)) is visible. 
The peak (b) is generated by a process occurring in the tagging MCP. A secondary electron is detected by the tagging MCP through an avalanche process where around $10^6$ electrons are generated, which impinge onto the anode with energies of several hundred \si{\electronvolt}. In this process, photons are created, which can then be detected by MCP2 \cite{2015_Khaw}. Peak (a) represents the true starting time and is most likely created by both MCPs very rarely picking up the same noise (e.g. sparks). The origin of peak (c) is not yet fully understood.

\subsection{\label{sec:foil_characterisation}Time Correction}

To accurately extract the characteristics of the foil, the TOF must be corrected to the true starting time, $t_\mathrm{corr}$. While the most straightforward approach would be to determine the time of peak (a), it is not always visible and, due to its low statistics, using it to determine the true $t_\mathrm{corr}$ would introduce a significant uncertainty. Instead, peak (b) is fitted  with a Langaus function (Landau convoluted with a Gaussian \cite{langaus}) to determine $t_0$. In addition to this, for the time correction we then need to account for (i) the time taken by the secondary electrons to travel from the foil to the tagging MCP ($t_{e^-}$ = \SI{11.5}{\nano\second} \cite{2015_Khaw}) and (ii) the time for the photon to travel from foil to the MCP2 (\SI{563}{\milli\meter} distance, $t_\gamma$ = \SI{1.9}{\nano\second}). Therefore, the full time correction is expressed as:
\begin{equation}
    t_\mathrm{corr} = t_0 - (t_{e^-}+t_{\gamma}),
\end{equation}
which agrees well with the time of the small peak (a).

\subsection{\label{sec:energy_characterisation}Energy Characterisation}
\begin{figure}[t!]    \centering
	    \includegraphics[width=1\columnwidth, trim={0 5 35 25},clip]{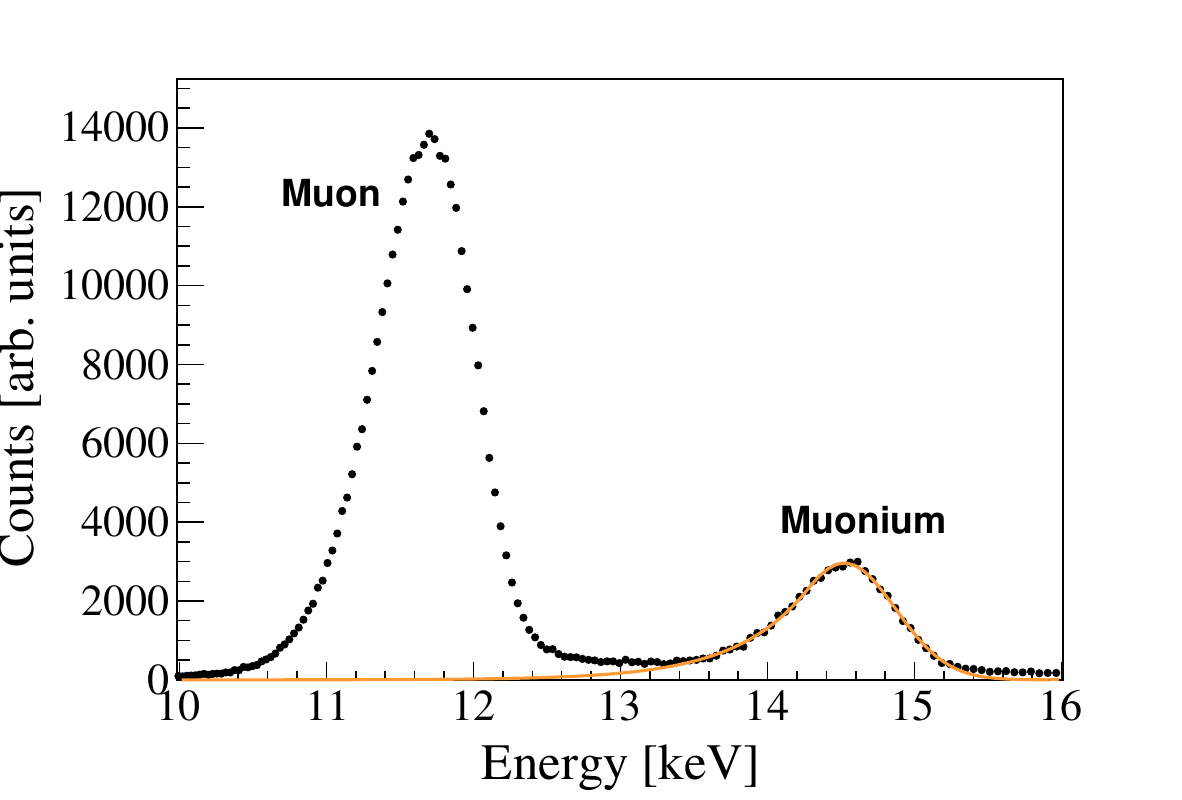}
	       \caption[]{\label{fig:M_Energy_12keV}The energy distribution, recorded with muons accelerated to \SI{12}{\kilo\electronvolt} at the moderator. The energy distribution for the muonium atoms is fitted with a Crystal Ball function (orange).} 
\end{figure}

To extract the energy characterisations of particles impinging on MCP2, we convert the corrected TOF for peaks (d) and (e) in Fig.\ref{fig:M_TOF_12keV} into an energy spectrum, as both the flight time and the distance are known. However, due to the tagging electric fields applied in the TC, the muon energy distribution is not unbiased for extracting the energy loss information. Instead, we fit a Crystal Ball function \cite{crystall_ball} to the hydrogen or muonium distribution (see Fig.~\ref{fig:M_Energy_12keV}). This allows us to extract both the width and the most-probable energy. The energy loss is then calculated as the difference between the incident and most-probable energy, while the width directly corresponds to the energy straggling in the carbon foil.

\begin{figure*}[t!]
    \centering
        \begin{subfigure}{0.99\columnwidth}
	  	\centering
	    \includegraphics[width=0.99\textwidth, trim={0 0 0 0},clip]{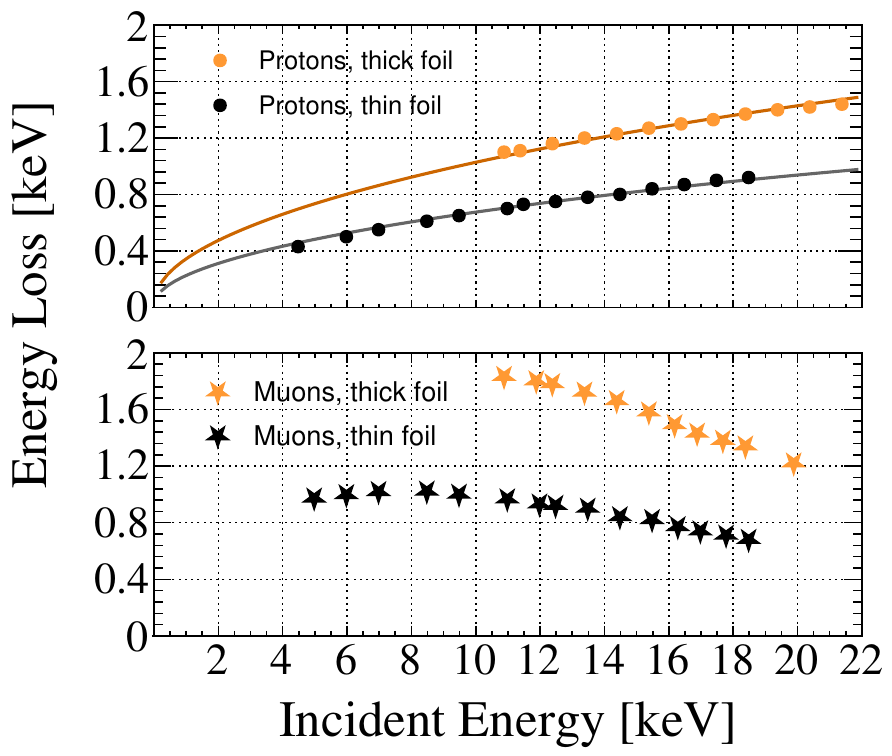}
	 \end{subfigure}
	 	\begin{subfigure}{0.99\columnwidth}
	    \centering
	    \includegraphics[width=0.99\textwidth,trim={0 0 0 0},clip]{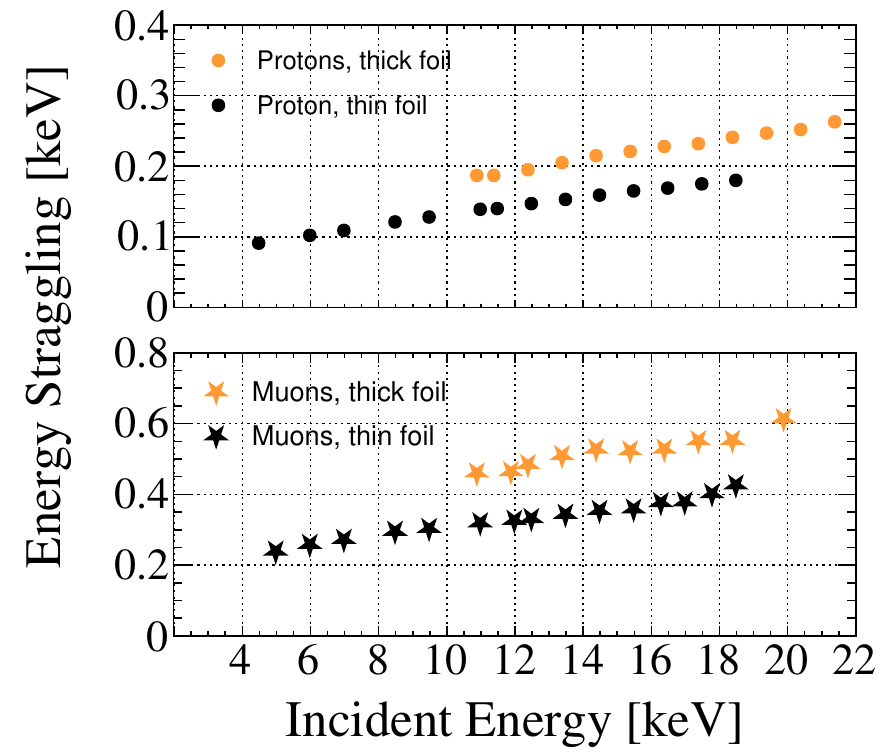}
	  \end{subfigure}

	  \caption[]{\label{fig:energy_loss} The energy loss and straggling of protons (full circle) and muons (stars) measured with the thick (\SI{2.6}{\micro\gram\per\centi\meter\squared}, orange) and thin (\SI{1.7}{\micro\gram\per\centi\meter\squared}, black) foil. The solid lines are fits of the energy loss of protons to Eq.~\ref{eq:energy_loss}~\cite{2015_Allegrini}.}
\end{figure*}

Ultra-thin carbon foils at various thicknesses were previously studied by Allegrini \emph{et al.} with protons \cite{2015_Allegrini}. These were used to formulate a semi-empirical description of the energy loss, given as:
\begin{equation}
\label{eq:energy_loss}
\Delta E = k \frac{Nd}{a_0 + a_1 E_0^{-0.4} + a_2 E_0^{0.25}},
\end{equation}
where $\Delta E$ represents the energy loss, $E_0$ is the kinetic energy before interaction with the foil (incident energy), $k$ is a unit conversion factor (= \num{19.9}), $a_0$, $a_1$ and $a_2$ are fitting parameters found from their data and tabulated in Ref.~\cite{2015_Allegrini}, and $Nd$ is the areal foil thickness. Eq.~(\ref{eq:energy_loss}) was used to fit our data where we fixed the value of $a_0$, $a_1$ and $a_2$ to those reported in Ref.~\cite{2015_Allegrini} but treated $Nd$ as a free parameter to extract the ``effective" foil thickness. From these fits we find that for the proton data with the thick foil, we extracted a thickness of \SI{2.59+-0.01}{\micro\gram\per\centi\meter\squared}, which significantly deviates from the nominal value of \SI{2.00}{\micro\gram\per\centi\meter\squared} based on the foil specifications. Similarly, for the thin foil proton data, we obtained a thickness of \SI{1.70+-0.01}{\micro\gram\per\centi\meter\squared} instead of the nominal value of \SI{0.5}{\micro\gram\per\centi\meter\squared}. Similar deviations were also observed by Allegrini \emph{et al.} Nonetheless, this smaller-than-expected reduction in thickness already has a significant impact on the low-energy muons. The energy loss is roughly cut to half, the straggling is significantly reduced by around \SI{30}{\percent}, and the energy distribution is more symmetric. The results for energy loss and straggling are summarized in Fig.~\ref{fig:energy_loss}.

\subsection{\label{sec:neutrals}Neutral Formation}

Another interesting performance aspect of a carbon foil is the fraction of incident charged particles that leave the foil as neutral atoms. In the case of protons, hydrogen (H) emerges from the foil, while for $\mu^+$ it forms muonium (Mu). To calculate the neutral formation efficiency, we determine the relative amount of neutrals compared to the total amount of neutrals and charged particles. In the TOF spectrum, the hydrogen or muonium distribution is fitted with a Langaus function (see Fig.~\ref{fig:M_TOF_12keV} peak (d) for muonium and peak (e) for muons). Due to the electrical fields required in the tagging setup, the charged particles experience a slight deflection compared to neutral atoms, decreasing their transport efficiency to MCP2. To adjust for these losses, we utilized a detailed SIMION \cite{SIMION} simulation of the LEM beamline.

\begin{figure}[t!]
    \centering
	    \includegraphics[width=1\columnwidth, trim={0 0 0 0},clip]{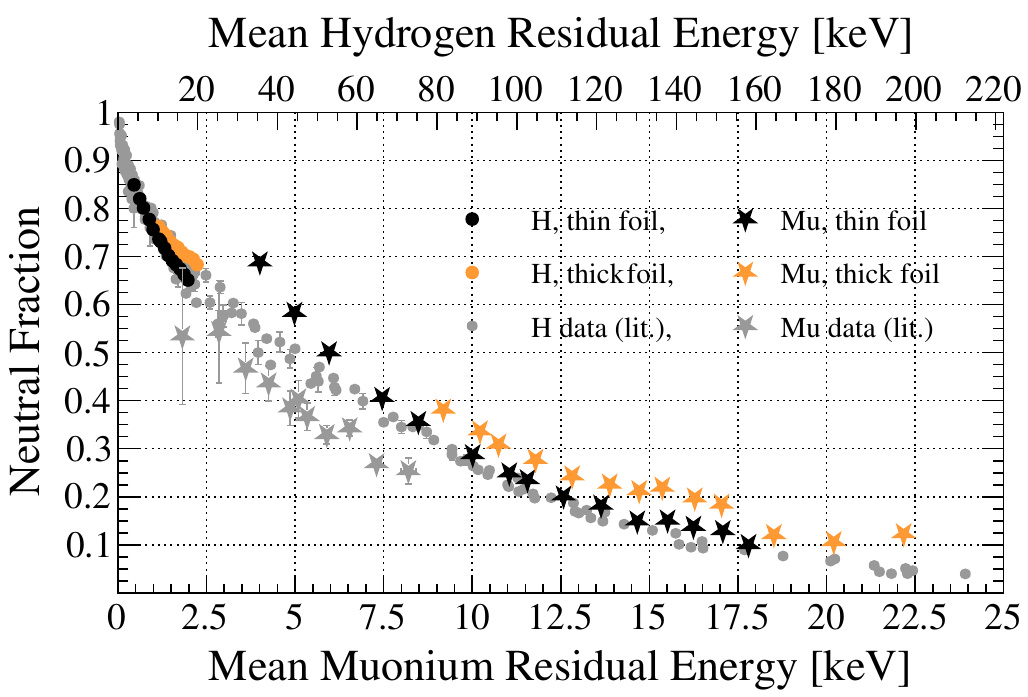}
	       \caption[]{\label{fig:neutral_efficiency}The neutral formation efficiency of hydrogen (full circle) and muonium (star) shown for the thick foil (orange) and the thin foil (black). The literature data (gray) for hydrogen was taken from \cite{1972_Berkner, 1973_Meggitt, 1982_Kreussler, 1987_Mazuy, 1993_Kallenbach,1992_Funsten, 2014_Allegrini, 2016_Allegrini}, the muonium data is taken from the Mu-MASS collaboration \cite{2020_Janka}.}
\end{figure}

While neutral atoms formation efficiency is disadvantageous for \muSR\ experiments as it reduces the polarization of the beam reaching the sample, it holds high significance for precision spectroscopy experiments requiring muonium beams, such as the measurement of the Lamb shift \cite{2021_Ohayon,2022_Janka}. The results of neutral formation are summarized in Fig.~\ref{fig:neutral_efficiency} and show that data for hydrogen from both the thick and thin foils agree well with data taken from the literature \cite{1972_Berkner, 1973_Meggitt, 1982_Kreussler, 1987_Mazuy, 1993_Kallenbach,1992_Funsten, 2014_Allegrini, 2016_Allegrini}. To enable comparison with experiments involving different foil thicknesses, the data is plotted for the most-probable energy the particles have when leaving the foil (mean residual energy), as neutral formation predominantly occurs close to the exit surface. In the case of muonium, the thin foil data at higher energies follows the trend of the hydrogen formation curve. However, at lower energies, it exhibits a slight increase. The muonium data with the thick foil suggests that the formation efficiency is even slightly higher than that of the thin foil. Notably, the efficiencies from both datasets reported here are significantly higher than what was seen from the sparse literature data \cite{2020_Janka}.

\subsection{\label{sec:tagging}Tagging Efficiency}

The tagging efficiency is derived from a single dataset where the position-sensitive MCP2 acts as a trigger. The total amount of hits detected by the MCP2 is used for normalization, while the coincidence events with the tagging MCP represent the number of tagged particles detected. The tagging efficiency is then calculated as the ratio between these two values. 

An analysis of the tagging efficiency extracted from the proton data reveals its energy dependence. A comparison of proton data indicates that the thin foil achieves a slightly higher tagging efficiency. This suggests that it is more uniform, homogeneous, and with fewer holes for the particles to pass through without interaction. Interestingly, for muon energies typically used in \muSR\ measurements (\SIrange{13.5}{18.5}{\kilo\electronvolt}), the tagging efficiency for muons is higher compared to protons. Additionally, it remains nearly constant and only starts to drop at lower energies.

Allegrini \emph{et al.} observed in their hydrogen data that foil contamination can increase both the tagging efficiency and neutral formation to some extent \cite{2016_Allegrini}, which could explain why a higher neutral formation was observed with the thick foil as mentioned earlier.

\begin{figure}[t!]    \centering
	    \includegraphics[width=1\columnwidth, trim={0 0 0 0},clip]{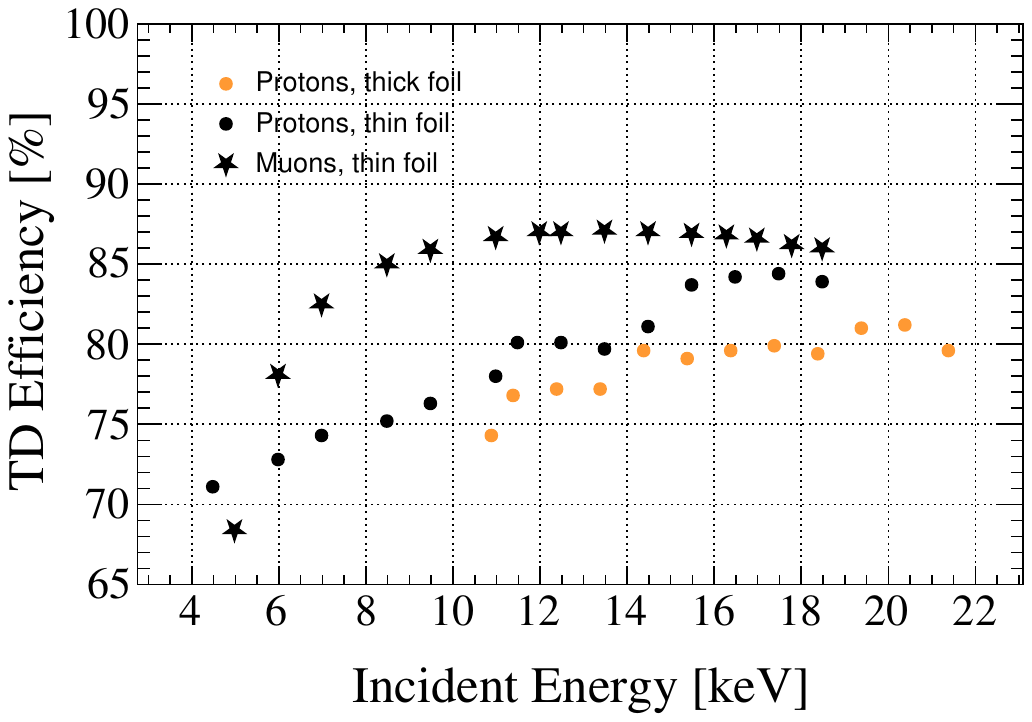}
	       \caption[]{\label{fig:tagging_efficiency}The tagging efficiency of protons (full circle) and muons (star) shown for the thick (orange) and the thin foil (black).}
\end{figure}

\subsection{\label{sec:beamspots}Achievable Beamspots}

To achieve the smallest beam spots, the voltages applied on the Einzel lens 3 and the ring anode were optimized. The beam spots are constructed by selectively choosing only TOFs from muons or protons and neglecting the neutral particles which can not be focused.
The figure-of-merit of the beam spot sizes was chosen to be the standard deviation of a 2D Gaussian, where $x$ denotes the horizontal and $y$ the vertical axis. 

\begin{figure}[b!]
    \centering
	    \includegraphics[width=1\columnwidth, trim={0 0 0 0},clip]{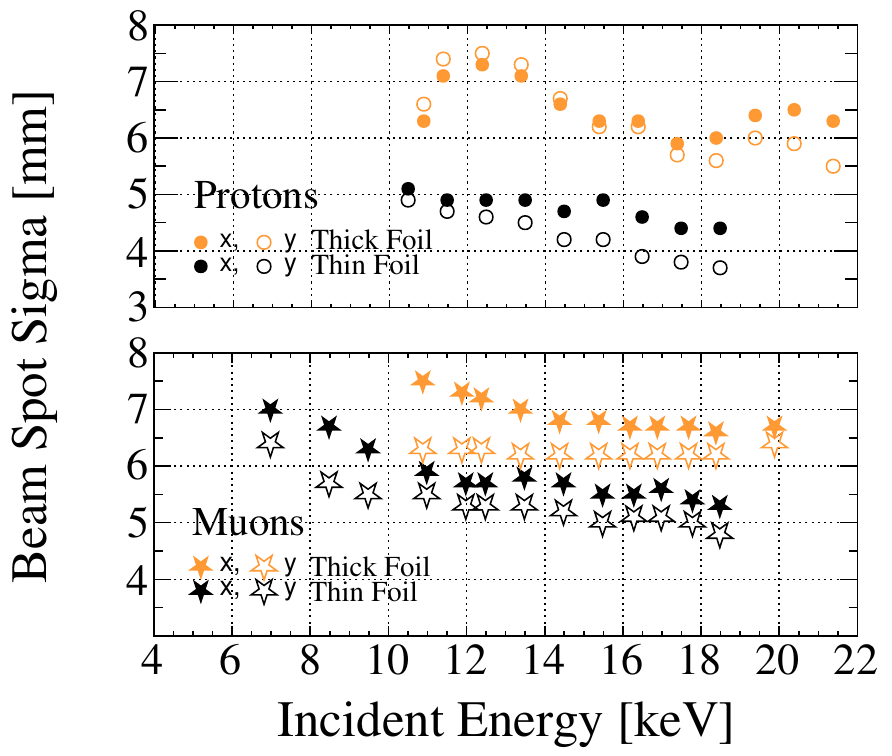}
	       \caption[]{\label{fig:beamspot}The smallest achievable beam spot sigmas in horizontal ($x$) and vertical ($y$) direction for protons (full circle) and muons (star) shown for the thick (orange) and the thin foil (black).}
\end{figure}

In Fig.~\ref{fig:beamspot} the smallest beam spot sizes are summarized for both protons and $\mu^+$. As expected from the smaller energy loss and straggling for protons compared to muons shown in Fig.~\ref{fig:energy_loss}, also the achievable beam spot size follows that trend. 
Comparing the proton data from the thick foil with the thin one, there are some challenges in making a direct comparison since the beam parameters were optimized differently. In the data with the thick foil, only for a few energies the high voltage settings on the different lenses were optimized via measurements; for the others the settings were scaled by energy. This assumption may not be accurate enough for the moderator settings for protons, as the initial conditions of the protons, such as the position where they are emitted, are not well known. In contrast, in the case of the thin foil data, all electrical fields were scanned and optimized to get the best beam spot. Therefore, a large improvement is noticeable, and the non-monotonic variations vanish.

For $\mu^+$, both datasets were optimized in the same manner, making them comparable and relevant to all measurements performed at LEM. At the energies of interest for \muSR\ measurements in the range of \SIrange{13.5}{18.5}{\kilo\electronvolt}, the thin foil allows for a beam spot roughly \SI{1}{\milli\meter} ($\sim 15\%$) smaller in sigma with the new foil. 

Interestingly, the data suggests that below \SI{10}{\kilo\electronvolt} the beam spot sigmas in the horizontal direction $x$ improves by more than \SI{1}{\milli\meter}. The gradual increase of beam spot size in $x$ direction appears below around \SI{14}{\kilo\electronvolt} for the thick foil data, while it is observed only at energies below \SI{11}{\kilo\electronvolt} for the thin foil data. These results are promising for experiments such as Mu-MASS \cite{2018_Crivelli}, where a combination of sub-\SI{10}{\kilo\electronvolt} $\mu^+$ with the smallest possible beam spot size is beneficial. 

The origin of the systematically larger beam spot sigma in horizontal direction $x$ is attributed to the moderator extraction. Since the moderated muons need to pass through three vertical grids, they experience forces mainly in the horizontal direction, distorting the phase space, while in vertical direction they remain unaffected.

\section{\label{sec:aging}Foil Aging and Cleaning}

The aging effect of the foil is predominantly observed as an increasing thickness with time due to contaminants such as water or other out-gassing from samples in the vacuum. To study this effect, proton datasets were acquired at various points in time and were used to determine the energy loss and hence the ``effective" foil thickness. The first dataset was taken directly after installation of the foil, which resulted in a thickness of \SI{1.90+-0.01}{\micro\gram\per\centi\meter\squared}. After two months in vacuum at \SI{1e-9}{\milli\bar} a second dataset was recorded, from which we observe an increased thickness of \SI{1.96+-0.01}{\micro\gram\per\centi\meter\squared}. This is attributed to water adsorbed to the surface of the foil. The straggling and neutral formation however did not change significantly, while the tagging efficiency decreased slightly. 
\begin{figure*}[t!]
    \centering
            \begin{subfigure}{0.99\columnwidth}
	  	\centering
	    \includegraphics[width=0.99\textwidth, trim={0 0 0 0},clip]{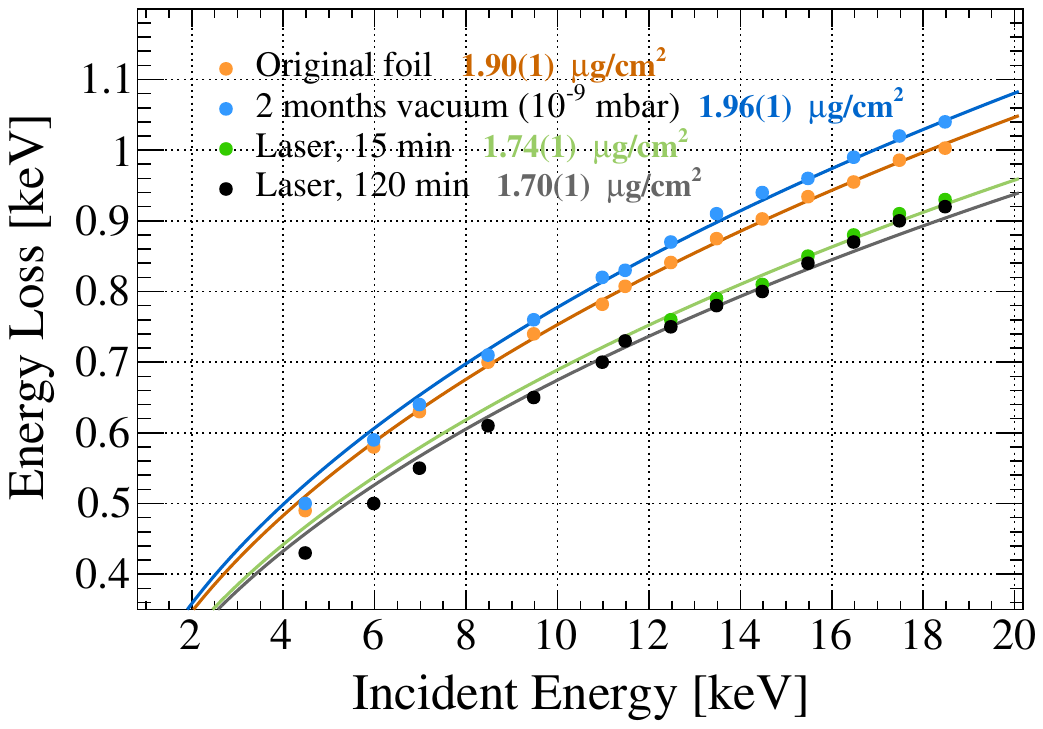}
	 \end{subfigure}
	 	\begin{subfigure}{0.99\columnwidth}
	    \centering
	    \includegraphics[width=0.99\textwidth,trim={0 0 0 0},clip]{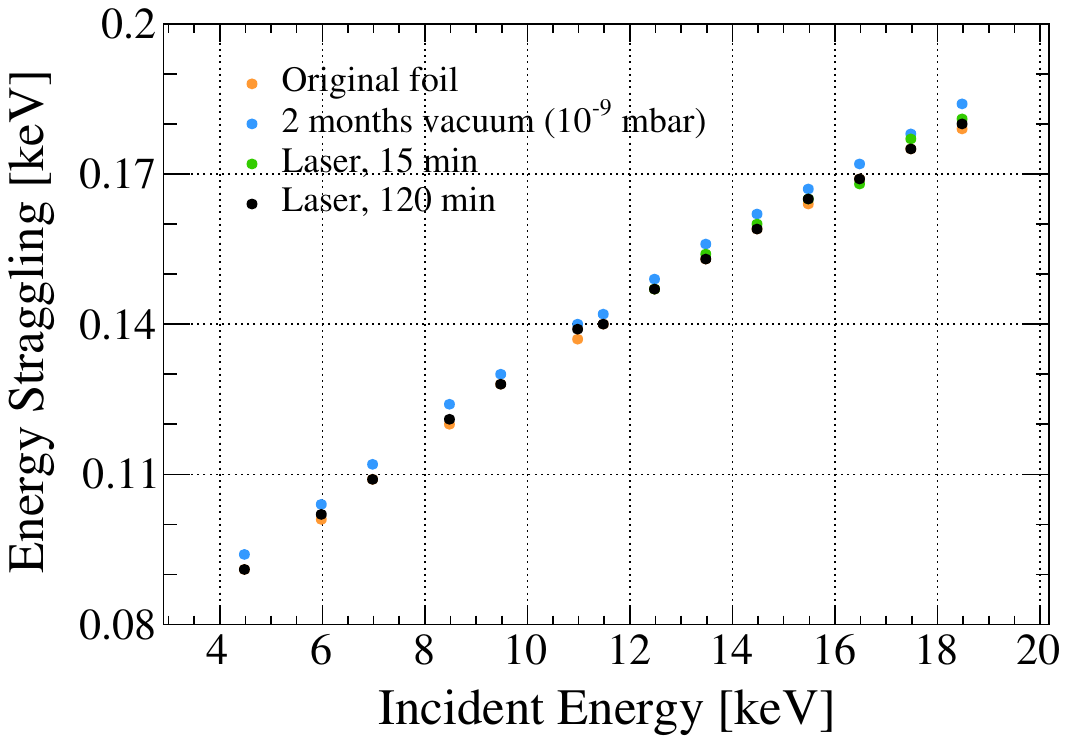}
	  \end{subfigure}
        \begin{subfigure}{0.99\columnwidth}
	  	\centering
	    \includegraphics[width=0.99\textwidth, trim={0 0 0 0},clip]{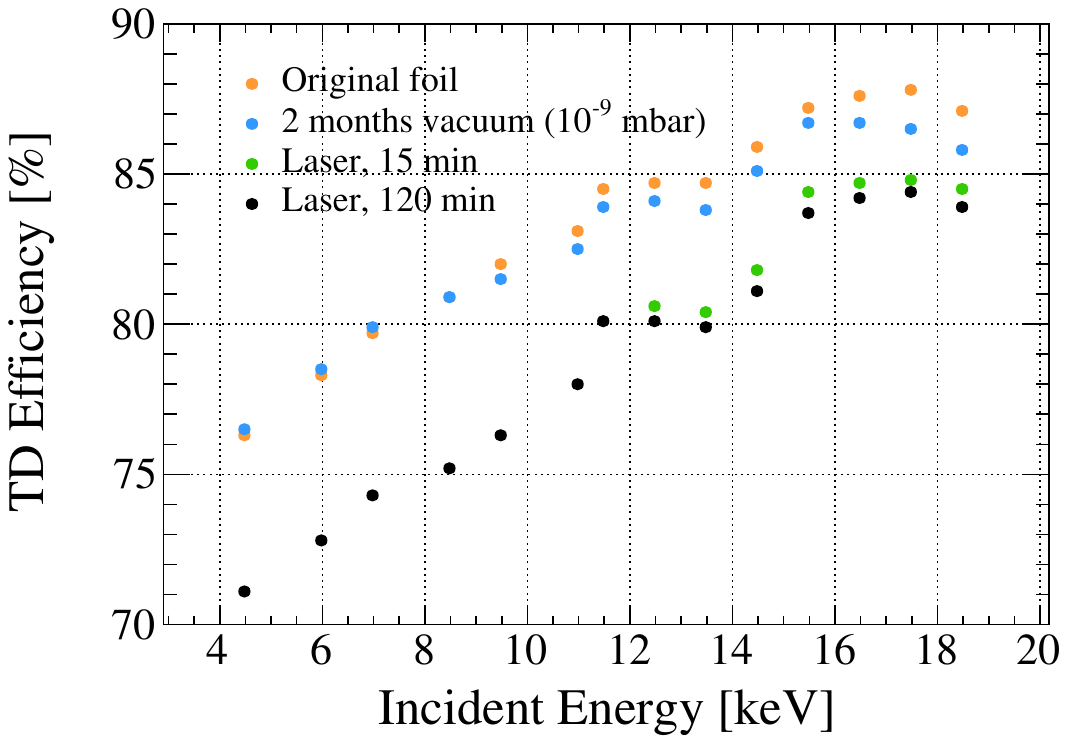}
	 \end{subfigure}
	 	\begin{subfigure}{0.99\columnwidth}
	    \centering
	    \includegraphics[width=0.99\textwidth,trim={0 0 0 0},clip]{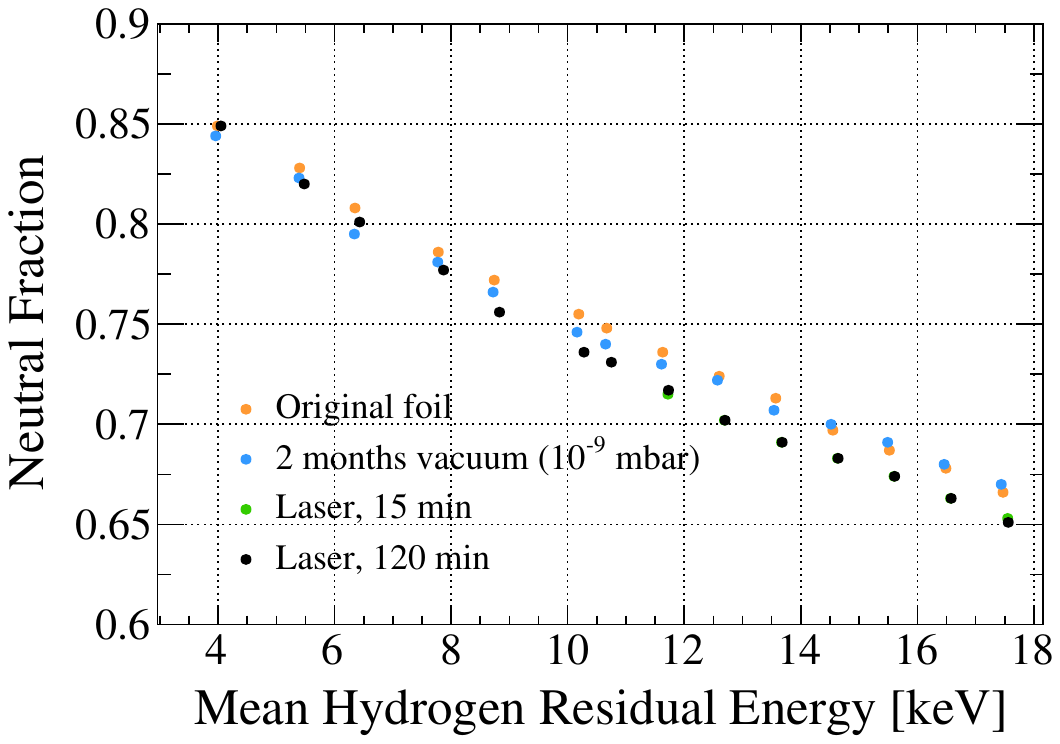}
	  \end{subfigure}

	  \caption[]{\label{fig:td_eff_cleaning}The energy loss (top left) and straggling (top right), the tagging efficiency of protons (bottom left) and the hydrogen formation probability (bottom right) measured with a nominally \SI{0.5}{\micro\gram\per\centi\meter\squared} foil. The data points were taken at different times, before (orange, blue) and after (green, black) illumination with blue laser light. The solid lines are fits of the energy loss of protons to Eq.~\ref{eq:energy_loss}~\cite{2015_Allegrini}.}
\end{figure*}

In an effort to remove the water layers from the foil, a continuous wave blue laser (\SI{457}{\nano\meter}, MBL-W-457nm-3W, Del Mar Photonics \cite{DelMarPhotonics}) was positioned at an optical port near the mirror location (see Fig.~\ref{fig:lem}). A lens system was used to defocus the laser beam to homogeneously illuminate the foil and prevent burning a hole into it, while gradually heating it. The output power of the laser was set to reach approximately \SI{100}{\milli\watt\per\centi\meter\squared}. Simultaneously, the tagging MCP was kept active to record and monitor the amount of charged particles being released from the foil.
During illumination, the count rate recorded by the tagging MCP increased from a few tens of hertz (dark counts) to several tens of kilohertz, then it decreased exponentially with time. After illumination for \SI{15}{\minute}, the tagging MCP rate reached a plateau of around one kilohertz. Following this foil cleaning procedure we measured the energy loss again, which revealed that its thickness decreased from \SI{1.96+-0.01}{\micro\gram\per\centi\meter\squared} to \SI{1.74+-0.01}{\micro\gram\per\centi\meter\squared}, i.e. even thinner than the initial ``as delivered" thickness. The tagging efficiency reduced by around \SI{5}{\percent}, and the neutral formation efficiency at energies larger than \SI{8}{\kilo\electronvolt} decreased by \SI{3}{\percent}. 
Longer laser illumination for an additional \SI{2}{\hour} resulted in a further thickness reduction to \SI{1.70+-0.01}{\micro\gram\per\centi\meter\squared}, but it eventually stabilized. This additional improvement did not have a significant impact on straggling, tagging efficiency, or neutral formation efficiency. This thickness measurement agrees very well with what Allegrini \emph{et al.} found for their nominally \SI{0.5}{\micro\gram\per\centi\meter\squared} foil \cite{2015_Allegrini}. We used this cleaned foil for all the measurements shown in Sec.~\ref{sec:foil}. A summary of the measurements before and after laser treatment is presented in Fig.~\ref{fig:td_eff_cleaning}.

\section{\label{sec:summaryn}Summary and Outlook}

The replacement of the nominally \SI{2.0}{\micro\gram\per\centi\meter\squared} foil with a nominally \SI{0.5}{\micro\gram\per\centi\meter\squared} foil resulted in a significant improvement in the beam quality of the LEM beamline. This is despite the fact that the final thickness (\SI{1.7}{\micro\gram\per\centi\meter\squared}) is larger than initially expected. Particularly noteworthy are the improvements in beam spot size, energy loss, and energy straggling. Moreover, the introduced method for cleaning the carbon foil from contamination via laser illumination holds the promise of providing more controlled and reproducible conditions for all measurements conducted at LEM. These encouraging results provide motivation for future efforts aimed at further reducing the foil thickness.

The potential for using even thinner carbon foils, e.g. $\approx$ \SI{2}{\nano\meter} thick, or even less by transitioning to single or a few layers of graphene ($\approx$ \SI{0.3}{\nano\meter} thick per layer) presents an exciting prospect. Studies show that the secondary electron emission and therefore the tagging efficiency does not decrease significantly when reducing the foil thickness or changing to graphene \cite{2016_Allegrini}, and therefore would make it a very promising candidate for implementation at LEM. However, while such reductions have been achieved for smaller foil areas (e.g.~\SI{3}{\milli\meter} diameter circle) and with less transparent grids (\SI{36}{\percent})\cite{2020_Vira}, implementing these changes for a larger area such as \SI{40}{\milli\meter} diameter circle with a more than \SI{90}{\percent} transparent grid will pose a big challenge.

Future developments at LEM will involve a re-evaluation of the use of grids and their wire density. 
The current tagging setup, which requires the muons to pass through five grids, has a total transmission of roughly \SI{80}{\percent}, significantly affecting the rate of muons reaching the sample and the minimal achievable beam spot size.
Similar considerations can be made at the moderator position. The moderator extraction fields are applied through three \SI{95}{\percent} transparent vertical grids, resulting in a combined transparency of roughly \SI{90}{\percent}. By changing the moderator extraction to a SOA lens \cite{1986_SOA}, the number of grids at the moderator could be reduced to one. This change also provides an opportunity to address the asymmetry between horizontal and vertical beam spot sigma.

\begin{acknowledgments}
All the measurements have been performed at the Swiss Muon Source S$\mu$S, Paul Scherrer Institute, Villigen, Switzerland. This work is supported by the Swiss National Science Foundation under the grants 192218 (MM) and 220823 (GJ).
\end{acknowledgments}


\bibliography{apssamp}

\end{document}